\documentclass[a4paper,11pt]{article}
\usepackage{pos}
\usepackage{floatrow}
\usepackage{bm}

\newcommand{\norm}[1]{\left\lVert#1\right\rVert}

\title{Sensitivity of the Cherenkov Telescope Array to a dark matter signal from the Galactic centre}
 \ShortTitle{CTA: DM in the GC}

\author*[a]{Christopher Eckner}

\affiliation[a]{Univ.~Grenoble Alpes, USMB, CNRS, LAPTh, F-74000 Annecy, France}

\forColl{CTA consortium} 

\emailAdd{eckner@lapth.cnrs.fr}

\abstract{
High-energy gamma rays are promising tools to constrain or reveal the nature of dark matter, in particular Weakly Interacting Massive Particles. Being well into its pre-construction phase, the Cherenkov Telescope Array (CTA) will soon probe the sky in the 20 GeV - 300 TeV energy range. Thanks to its improved energy and angular resolutions as well as significantly larger effective area when compared to the current generation of Cherenkov telescopes, CTA is expected to probe heavier dark matter, with unprecedented sensitivity, reaching the thermal annihilation cross-section at ~1 TeV.

This talk will summarise the planned dark matter search strategies with CTA, focusing on the signal from the Galactic centre. As observed with the Fermi LAT at lower energies, this region is rather complex and CTA will be the first ground-based observatory sensitive to the large scale diffuse astrophysical emission from that region. We report on the collaboration effort to study the impact of such extended astrophysical backgrounds on the dark matter search, based on Fermi-LAT data in order to guide our observational strategies, taking into account various sources of systematic uncertainty.

}

\FullConference{37$^{\rm{th}}$ International Cosmic Ray Conference (ICRC 2021)\\
		July 12th -- 23rd, 2021\\
		Online -- Berlin, Germany}

\begin{document}
\maketitle

\section{Introduction}
\label{sec:intro}

The cosmological model of a universe filled with cold dark matter (CDM), dark energy ($\Lambda$) and baryonic matter whose dynamics is governed by the laws of General relativity has a strong predictive power. Observations on a huge variety of length scales have shown that this paradigm is capable of explaining, for example, the shape of the cosmic microwave background's anisotropy power spectrum, the formation of large-scale structure with its particular filaments and voids as well as the dynamics of galaxy clusters and spiral galaxies \cite{Bertone:2004pz, Bertone:2016nfn}. Despite problems on, most noteably, small scales -- that may be ascribed to the insufficient understanding of feedback effects among different types of matter \cite{Bullock:2017xww} -- the $\Lambda$CDM model is the current benchmark theory of cosmology \cite{Aghanim:2018eyx}. 

Nonetheless, the question of the nature of DM still remains an open one. It is even true that during the more than 40 years of intensive research on the nature of DM none of the darling candidates like `Weakly Interacting Massive Particles' (WIMPs), sterile neutrinos or axions have left a clear trace of their existence \cite{Bertone:2018krk} besides gravitational attraction. The upcoming decade, the 20s of the 21th century, will enable the scientific community to profit from an ever-increasing wealth of new experiments and devices that outshine their predecessors in terms of sensitivity, accuracy and precision. Among those intruments, the Cherenkov Telescope Array (CTA) will open the door to in-depth studies of Galactic and extragalactic astrophysical processes that generate very-high-energy gamma-ray emission in the TeV energy range.

While the WIMP DM mass range $<100$ GeV has been explored by the current gamma-ray experiments, the  CTA will offer a glimpse on the TeV sky with unprecedented sensitivity, and explore the multi-TeV DM parameter space. The most stringent constrains are derived from observations of the centre of the Milky Way, which in the CDM paradigm is expected to be the region of highest DM density in the `neighbourhood' of the Earth. This region is hence a suitable target to study the gamma-ray emission originating from WIMP DM pair-annihilation events. 

CTA is expected to reach a sensitivity where a large portion of the WIMP hypothesis, i.e.~thermally produced DM particles annihilating with a velocity-weighted cross-section of $\langle\sigma v\rangle\sim3\cdot10^{-26}\,\mathrm{cm}^3\mathrm{s}^{-1}$, becomes experimentally accessible and, thus, concretely falsifiable. However, this unprecedented sensitivity to TeV gamma-ray emission poses potentially new problems to DM searches in the Galactic centre (GC). At this point, it is not clear which astrophysical processes will be encountered and what will be their general impact on searches for DM annihilation products. In this proceedings article, we summarise the CTA consortium's effort to provide the best possible characterisation -- taking into account the current theoretical and experimental limitations on the astrophysical and instrumental part -- of the telescope's sensitivity to a DM signal from the Galactic centre. The fully comprehensive analysis can be found in \cite{Acharyya:2020sbj}. 

\section{A model for the $\gamma$-ray emission in the Galactic centre at TeV energies}
\label{sec:astro_background}

\begin{figure}[t!]
\centering
\floatbox[{\capbeside\thisfloatsetup{capbesideposition={right,top},capbesidewidth=5cm}}]{figure}[\FBwidth]
{\caption{
Summary of background and signal components considered in this work. These panels show the result of the convolution of a physical model with CTA's instrument response function generated with {\sf ctools}. The simulation is based on the Galactic centre survey campaign as described in Sec.~\ref{sec:analysis} focusing on the energy range from 100 to 500 GeV. \label{fig:template_collection}} }
{\includegraphics[width=1.01\linewidth]{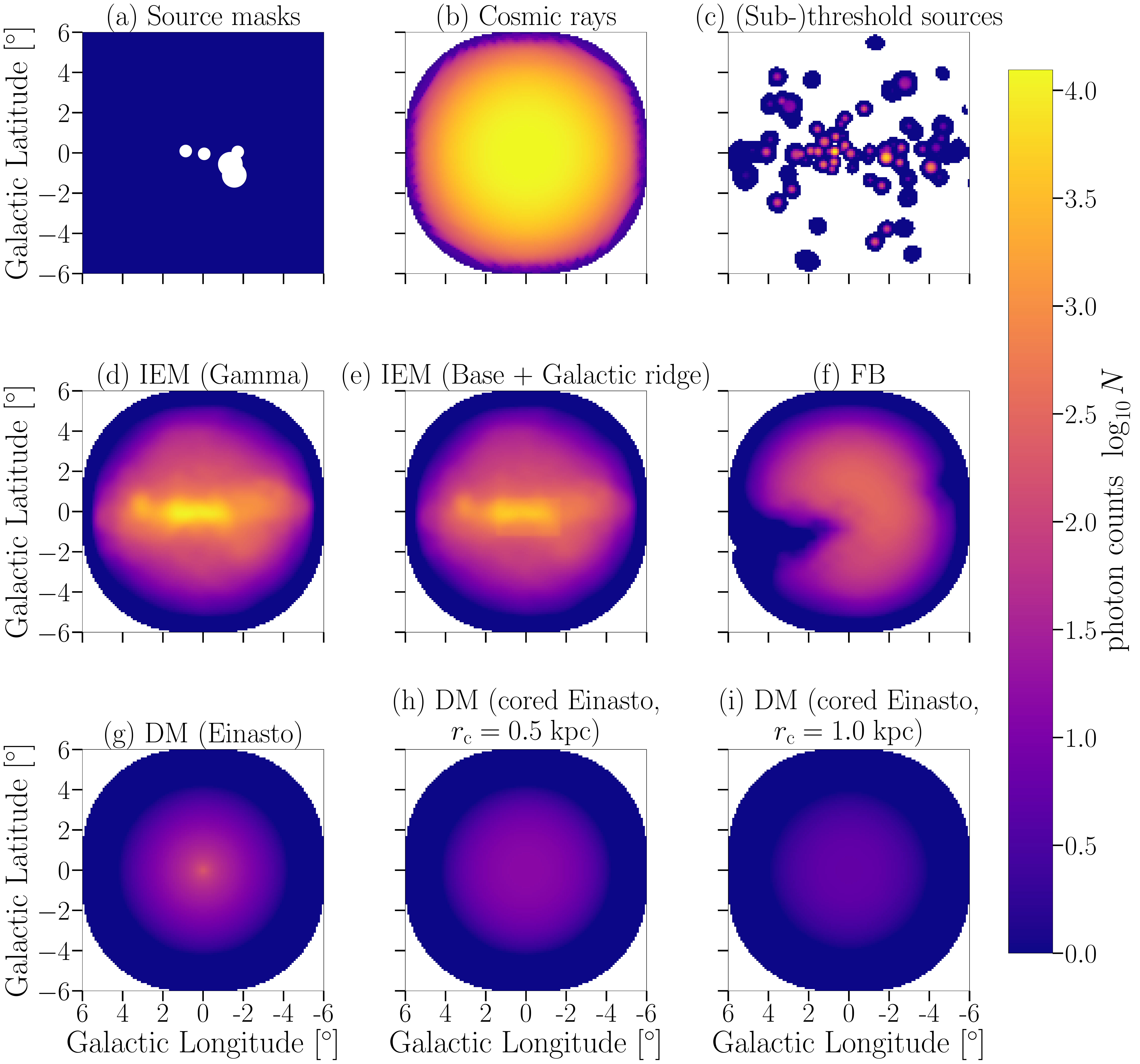}}
\end{figure}

The Milky Way's Galactic centre region hosts a plethora of individual, localised gamma-ray emitters as well as gas clouds, dust and photon radiation fields, which all may contribute in one or another way to the total gamma-ray luminosity of this part of the sky. The complexity of the prevalent emission processes makes it notoriously hard to uncover a DM signal from the astrophysical background --  most prominently exemplified by the ongoing, decade-long discussion about the so-called `Galactic centre GeV excess' discovered in the gamma-ray data taken by the Fermi satellite \cite{2009arXiv0910.2998G}. We base our model building endeavours on the learnings from current-generation gamma-ray telescopes like Fermi-LAT, MAGIC, VERITAS and H.E.S.S. regarding the components that are guaranteed contributions to the TeV gamma-ray sky. A summary of the considered components -- including instrumental background, astrophysical contributions as well as the DM component -- is shown in Fig.~\ref{fig:template_collection}.

\subsection{Irreducible cosmic-ray background}
\label{sec:IACT_background}

As a ground-based telescope, the CTA detects cosmic ray (CR) showers -- triggered by CRs that collide with particles in the Earth's atmosphere -- and from their properties it determines the nature of the primary incident particle (gamma ray or charged CR) as well as its energy and direction.
However, air showers from primary electrons and positrons are nearly indistinguishable from their gamma-ray counterparts while air showers of hadronic origin may be rejected at a much higher rate due to their distinctive detector signal. The CTA simulation working group has conducted a study to derive the expected morphology and spectrum of the instrumental background given the current layout design goals \cite{2017APh....93...76H}. We make use of their efforts by utilising the public code \emph{ctools}\footnote{http://cta.irap.omp.eu/ctools} that enables the user to automatically generate the instrumental background due to leptonic and hadronic cosmic rays for any pointing position of CTA including the convolution with the `Instrument Response Functions' (IRFs).

\subsection{Astrophysical emission components}
\label{sec:astro_background}

The available measurements of the gamma-ray luminosity of the GC region at energies below 100 GeV with Fermi LAT have revealed that this part of the sky is not substantially brighter than the remainder of the Galactic disc \cite{Ackermann:2012pya}. Current-generation ground-based Cherenkov telescopes like H.E.S.S., VERITAS and MAGIC have, in contrast, shown that in the TeV domain the GC is shining brighter than its surroundings \cite{Aharonian:2006au,Abdalla:2017xja,Archer:2016ein,Abramowski:2016mir}. We consider the following astrophysical emission components in this complex region of the Milky Way (see also Fig.~\ref{fig:template_collection}):
\begin{enumerate}
    \item [(i)] \emph{interstellar emission} (IE) being the result of interactions of primary cosmic rays with the the interstellar medium (gas, dust, radiation fields). We consider three models for the IE in the GC region. Two IE models, \emph{Gamma} and \emph{Base} discussed in \cite{Gaggero:2017jts}, where the Gamma model naturally explains and contains the bright TeV emission via a spatially dependent cosmic-ray diffusion coefficient whereas the Base model does not reflect this component so that we add it by hand with the parameters and morphology found in \cite{Aharonian:2006au}. The third model is the current Pass8 Fermi-LAT diffuse background model\footnote{\url{https://fermi.gsfc.nasa.gov/ssc/data/access/lat/BackgroundModels.html}} extrapolated to TeV energies.
    \item [(ii)] \emph{localised gamma-ray sources} already detected by current-generation instruments. There are in total six TeV gamma-ray emitters in the vicinity of the GC listed in the online TeV source catalogue TeVCAT~\cite{tevcat2}. One of them is likely a part of the Galactic ridge so that we neglect it. The remaining five sources are masked in our analysis pipeline taking into account their intrinsic extension and the angular resolution of CTA.
    \item [(iii)] \emph{Fermi Bubbles} (FBs), in particular their low-latitude part. They are a possibly faint large-scale diffuse component whose exact morphology is not well understood. We incorporate the low-latitude part of the FBs using the spatial template derived in \cite{TheFermi-LAT:2017vmf} together with the associated spectrum fitted with a log-parabola from the same work.
    \item [(iv)] \emph{sub-threshold gamma-ray sources} whose collective emission adds a diffuse glow to the Galactic disc emission. Since it is impossible to make definite statements about the structure of this component, we rely on the sky model developed for the CTA Galactic plane survey \cite{GPSPreliminary}.
\end{enumerate}

\subsection{Dark matter annihilation}
\label{sec:dm_signal}

The expected (prompt) differential gamma-ray flux $\textrm{d}\Phi_{\gamma}/\textrm{d}E_{\gamma}/\mathrm{d}\Omega$ due to DM pair-annihilation in the GC at the top of the Earth's atmosphere is given by (see, e.g., \cite{Bringmann:2012ez})
\begin{equation}
\label{DMflux}
  \frac{d\Phi_{\gamma}}{d\Omega\, dE_\gamma} (E_\gamma,\psi) = 
  \left(\vphantom{\frac{dN_\gamma^{f}}{dE_\gamma}\sum_f}\frac{1}{4\pi} \int_\mathrm{l.o.s}
  d\ell(\psi) \rho_\chi^2(\mathbf{r})\right) 
  \left({\frac{\langle\sigma v\rangle_\mathrm{ann}}{2S_\chi m_{\chi}^2} \sum_f
  B_f\frac{dN_\gamma^{f}}{dE_\gamma}}\right) \,,
\end{equation}
which requires assumptions about the DM particle properties and its interactions with Standard Model particles (term in second parenthesis) as well as the spatial distribution of DM $\rho_{\chi}$ in the Milky Way -- the so-called $J-$factor, the term in the first parenthesis.

As we are focusing on classical WIMP particles, we assume them to be Majorana fermions ($S_{\chi} = 1$), which annihilate into a single Standard Model particle species $f\in\{b\bar{b}, \tau^+\tau^-, W^+W^-\}$ exclusively, i.e.~$B_f = 1.0$. We thereby cover the broad range from soft to hard annihilation spectra $\mathrm{d}N_{\gamma}^f/\mathrm{d}E_{\gamma}$, whose numerical values for different DM masses $m_{\chi}$ we adopt from \href{http://www.marcocirelli.net/PPPC4DMID.html}{PPPC}~\cite{Cirelli:2010xx}.\\
Regarding the $J$-factor towards the GC, we intend to examine to contrasting cases: The possibility that the DM halo exhibits to prominent cusp in its very centre or that it features a shallow core-like structure in its central part. Both alternatives are consistent with current data \cite{Karukes:2019jwa}. As a representative of a cuspy DM density distribution, we select the Einasto profile $\rho_{\mathrm{Ein}}$ whereas the family of cored profiles is represented by a cored Einasto profile $\rho_{\mathrm{cEin}}$ defined as follows:
\begin{equation}
\rho_{{\rm Ein}}\!\left(r\right)=\rho_{s}\exp\!\left(-\frac{2}{\alpha}\left[\left(\frac{r}{r_{s}}\right)^{\alpha}-1\right]\right){\rm ,} \quad \rho_{{\rm cEin}}\!\left(r\right)=
\begin{cases}
\rho_{{\rm Ein}}\!\left(r_{c}\right) & \textrm{if}\;r\leq r_{c}\\
\rho_{{\rm Ein}}\!\left(r\vphantom{r_{c}}\right) & \textrm{if}\;r>r_{c}
\end{cases}\,\mathrm{,}
\end{equation}
where $\alpha=0.17$, $r_{s}=20$ kpc and $r_c\in\left[0.0, 1.0\right]$. These distributions are normalised such that $\rho_{\mathrm{Ein}}(r_{\odot}) = \rho_{\odot}$, where $r_{\odot} = 8.5$ kpc denotes the distance of the Sun from the GC and $\rho_{\odot} = 0.4$ GeV$/\mathrm{cm}^3$ is the local DM density at the Sun's position. We derive two-dimensional sky maps of the resulting $J-$factors via the public code \href{https://clumpy.gitlab.io/CLUMPY/}{CLUMPY} \cite{bonnivard2016clumpy}.

\section{Analysis framework}
\label{sec:analysis}

\noindent\textbf{Observational strategy.} Our analysis is based on simulations of the performance of CTA South and the corresponding IRF version \texttt{South\_z20\_average\_50h} of the \texttt{prod3b-v1} release. This publicly available IRF library contains all information necessary to convolve input physics models in units of flux with the effective area, point spread function and energy dispersion of the instrument. We implement the envisaged pointing pattern of the `Galactic centre survey' observation campaign (525h in total) as described in \cite{Acharya:2017ttl} to obtain three-dimensional templates of background and signal components. To study cored DM profiles, we add the exposure gained by the `extended Galactic centre survey' (300h in total), which is a follow-up campaign targeting the region up to a Galactic latitude of $10^{\circ}$ above the Galactic disc.

\noindent\textbf{Statistical framework.} We conduct a binned three-dimensional/template-based analysis to derive the sensitivity of CTA to a DM signal from the GC. We construct a framework, which is able to incorporate systematic uncertainties of the instrument as well as the chosen astrophysical background components. To this end, we employ the generalised Poisson log-likelihood function 
\begin{equation}
-2\ln{\mathcal{L}\!\left(\left.\bm{\mu_K}\right|\bm{n}\right)}= 
\min_{\bm{\Delta B}}\left\{\sum_{k=1}^{\mathcal{N}}\left[\vphantom{\int}n_k\ln{\left(\mu_K\right)_k} - \left(\mu_K\right)_k\right] 
- \frac{1}{2}\sum_{k,l=1}^{\mathcal{N}}\left[\vphantom{\int}\Delta B_k \left(K^{-1}\right)_{kl} \Delta B_l\right]\right\}\rm{,}
 \label{eq:loglikelihood_cov}
\end{equation}
where $\bm{\mu_K}$ denotes the model data and it is defined by
\begin{equation}
\left(\mu_K\right)_k\equiv \sum_X \mu_{k}^X+\Delta B_k + A^{\chi}\mu_{k}^{\chi}\label{eq:model_counts_cov}\,.
\end{equation}
while $\bm{n}$ refers to the mock data that we prepare from the pool of background templates in order to create a proxy for future CTA data. The index in the sums of Eq.~\ref{eq:loglikelihood_cov} runs over the pixel number of the flattened templates whereas $X$ runs over those background components that are part of the respective mock data set. With $\Delta B_k$ we denote the background perturbations, which are treated as nuisance parameters and hence profiled over in the log-likelihood function. The covariance matrix $K$ we control whether we include systematic effects or whether we take into consideration merely the statistical uncertainty. For example, spatial systematic uncertainties are described via a two-parameter model introducing a correlation length $\ell_{\mathrm{S}}$ and a fluctuation amplitude $\sigma_{\mathrm{S}}$:
\begin{equation}
\left(K_{\rm S}\right)_{jj^{\prime}} = \sigma_{{\rm S}}^2 \exp\!{\left(-\frac{1}{2}\frac{\norm{\vec{r}_j - \vec{r}_{j'}}^2}{\ell_{\rm S}^2}\right)}\rm{,}
\end{equation}
where $\vec{r}_j$ is the central position of the $j-$th spatial template bins 
in degrees of Galactic longitude and latitude, and we use the norm on the unit sphere for the distance between two spatial bins. The numerical evaluation of these quantities as well as the derivation of the upper limits on $A^{\chi}$ are performed with the public code \emph{swordfish} \cite{Edwards:2017mnf, Edwards:2017kqw}.

\noindent\textbf{Benchmark choices.} The results of the following section rely on these benchmark choices: $12^{\circ}\times12^{\circ}$ region of interest centred on the GC binned into square pixels of size $0.1^{\circ}$ as well as 54 energy bins from 30 GeV to 100 TeV whose boundaries correspond to the $2\sigma$ containment of the array's energy resolution\footnote{c.f.~\url{https://www.cta-observatory.org/science/cta-performance/\#1472563318157-d0191bc5-0280}}. Instrumental systematic errors are modelled with $\ell_{\mathrm{S}} = 0.1^{\circ}$ and an amplitude of $1\%$. If not stated otherwise, all figures show our constraints including these instrumental systematic uncertainties. Our benchmark mock data $\bm{n}$ consist of the CR and IE Gamma model templates.

\section{Projected CTA sensitivity to DM in the Galactic centre}
\label{sec:results}

In this section, we illustrate a selected number of results from the CTA consortium publication of the analysis presented in this proceedings article \cite{Acharyya:2020sbj}. We refer the reader to this document for more details and a broader discussion of various aspects of this study.

$\vphantom{d}$

\begin{figure}[t!]
\centering\includegraphics[width=0.49\linewidth]{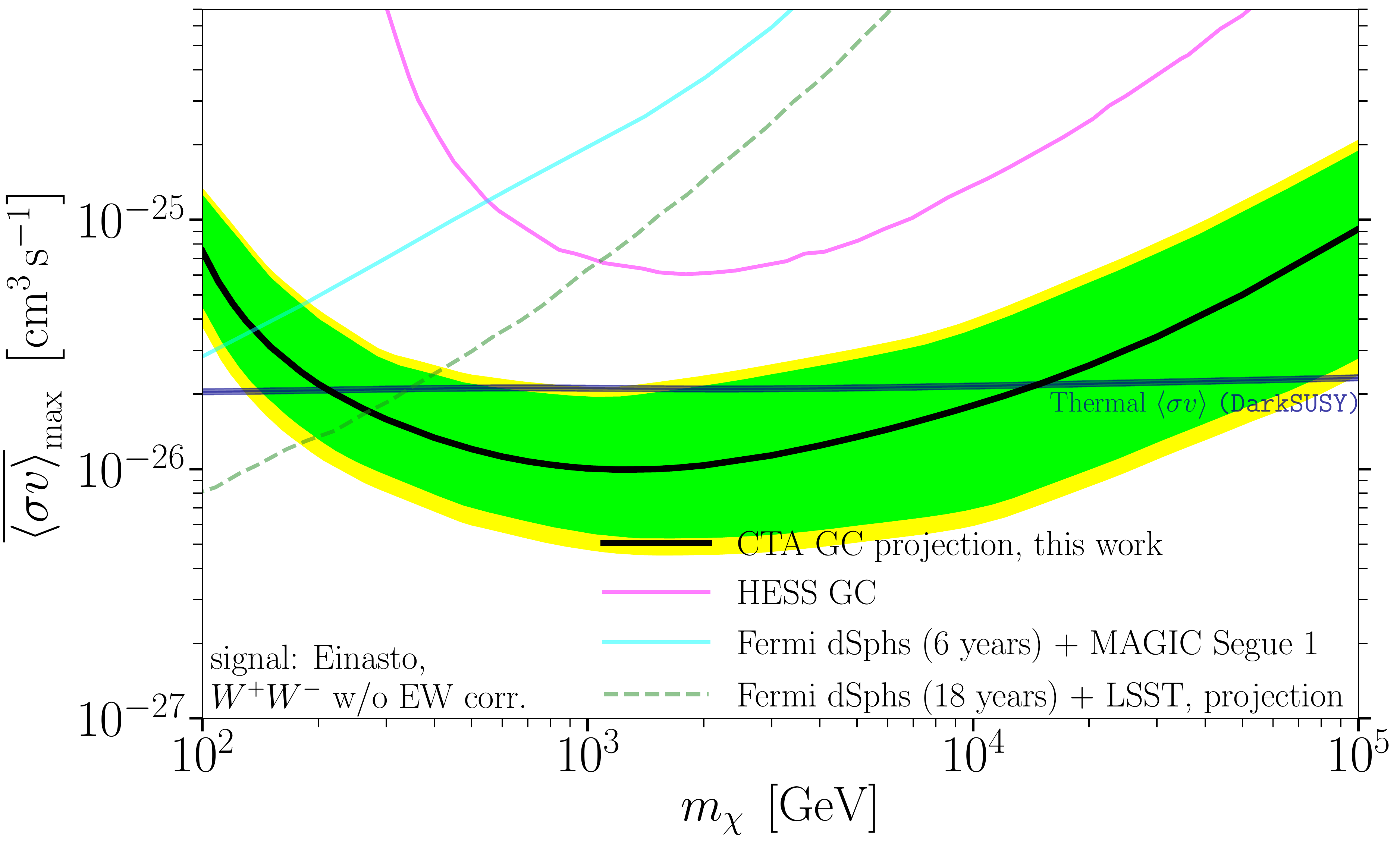}
\includegraphics[width=0.49\linewidth]{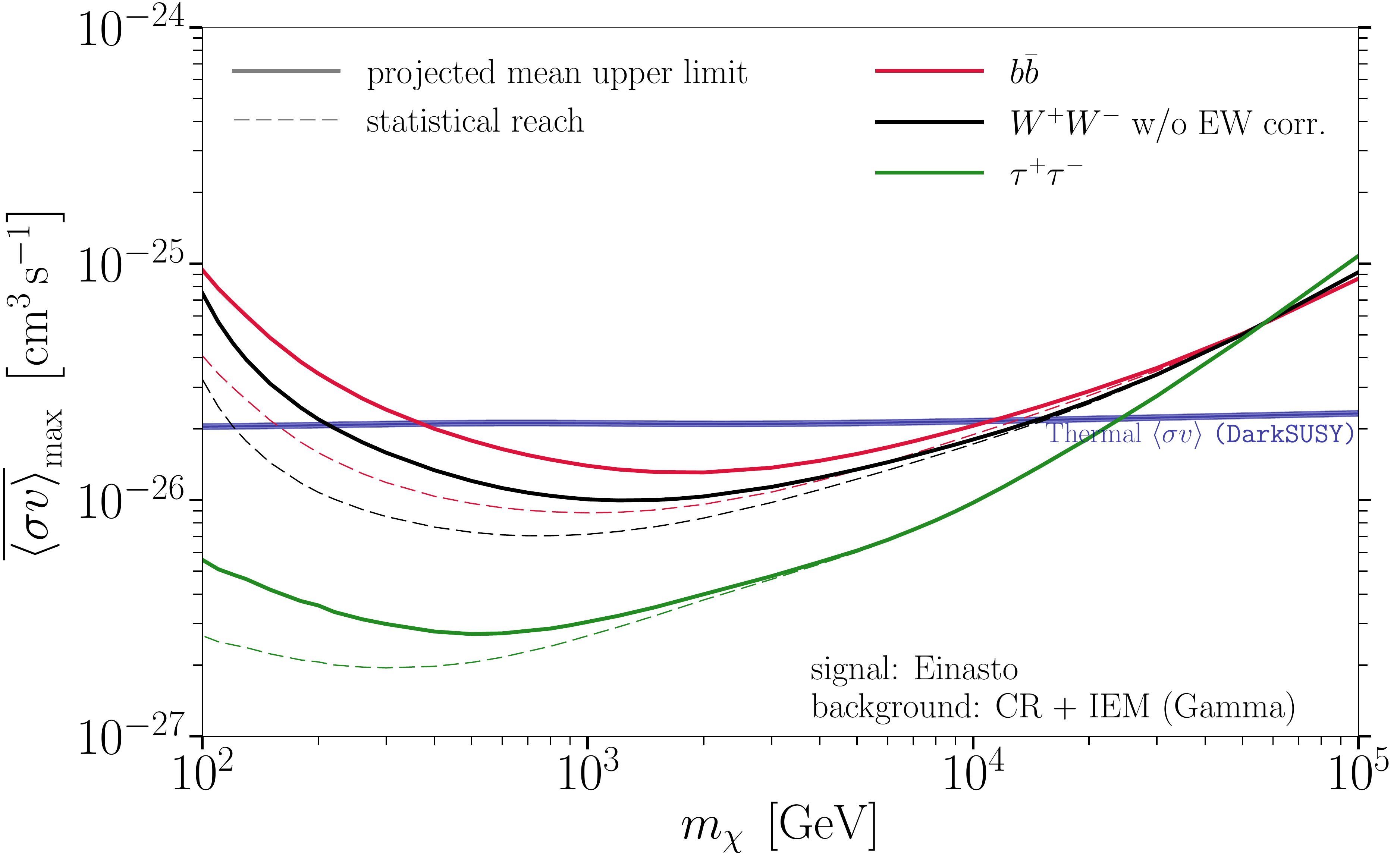}
\caption{Sensitivity of CTA to a DM annihilation signal assuming an Einasto profile, at 95\%\,C.L. in terms of the projected mean upper limits on the average velocity-weighted annihilation cross section, as  a function of the DM mass $m_\chi$. Results displayed with dashed lines lift the assumption of instrumental systematic errors. We highlight the cross-section needed to thermally produce DM in the early universe to match the cosmologically observed DM abundance (blue band), which we have calculated with \texttt{DarkSUSY} \cite{Bringmann:2018lay} and current Planck data \cite{Aghanim:2018eyx}.
\emph{Left panel:} Sensitivity to DM annihilation into $W^+W^-$ final states (black). The green (yellow) band indicates the $2\sigma$ ($3\sigma$) 
scatter of the projected limits (based on Monte Carlo realisations). For comparison, we display a set of constraints from current-generation instruments taken from \cite{Ackermann:2015zua, Abdallah:2016ygi, Drlica-Wagner:2019xan}.
\emph{Right panel:} DM annihilation into $\bar bb$ (red), $W^+W^-$ (black) and $\tau^+ \tau^-$ (green), respectively. \label{fig:bb_MCband}}
\end{figure}

\noindent\textbf{Sensitivity forecast for peaked DM profiles.} In Fig.~\ref{fig:bb_MCband} we summarise the projected upper limits on the DM annihilation signal for our analysis' benchmark choices. These constraints have been derived assuming an Einasto DM density profile. The right panel of this figure visualises that CTA is potentially capable of testing the thermal cross-section for a wide range of DM masses and annihilation channels. Moreover, the yellow (green) bands in the left panel exemplify that even the expected $2\sigma$ ($3\sigma$) scatter of these upper limits does not heavily alter this statement.

\begin{figure}[t]
\centering\includegraphics[width=0.6\linewidth]{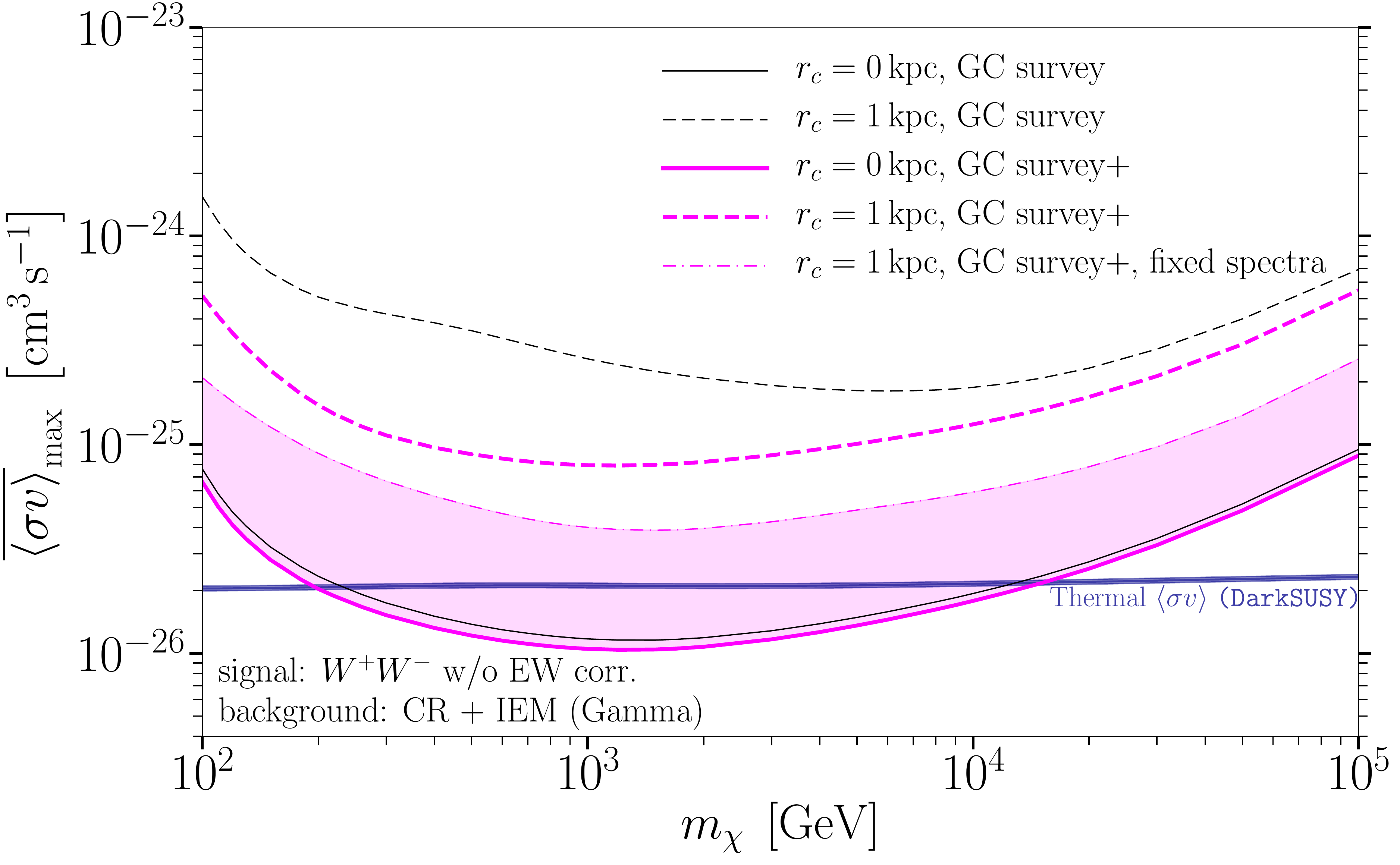}
\caption{CTA sensitivity to a DM signal, for the $W^+W^‐$ channel, comparing the case of an Einasto profile 
without core (solid) to that of an Einasto profile with a 1\,kpc core (dashed). 
Black lines show the sensitivity with the base survey only and magenta lines show the sensitivity from 
adding extended survey observations. Modelling the spectral information with greater care may lead to a 
further improvement of the sensitivity to a cored profile, as indicated by  the magenta dash-dotted line 
(see text for more details, including a discussion of the shaded area).
In order avoid excessive use of computational resources, the sensitivity predictions in this figure are 
based on only 20 (equally log-spaced) energy bins.
\label{fig:cored_templates_wwo_extended}} 
\end{figure}

\noindent\textbf{Sensitivity forecast for cored DM profiles.} Extensive DM cores in the GC region can be considered a `blind spot' of ground-based Cherenkov telescopes because the expected DM signal is degenerate with the bright CR instrumental background (see Fig.~\ref{fig:template_collection}). As shown in Fig.~\ref{fig:cored_templates_wwo_extended} this situation is ameliorated when we include the additional exposure from the extended Galactic centre survey (compare the black vs. the pink lines). In fact, these prospects can further be improved when the spectra of the background components are very well known (pink dot-dashed line). We simulate this situation by fixing the spectra of these components in our model data, i.e.~we introduce a large spectral correlation length.

\begin{figure}[t]
\centering\includegraphics[width=0.49\linewidth]{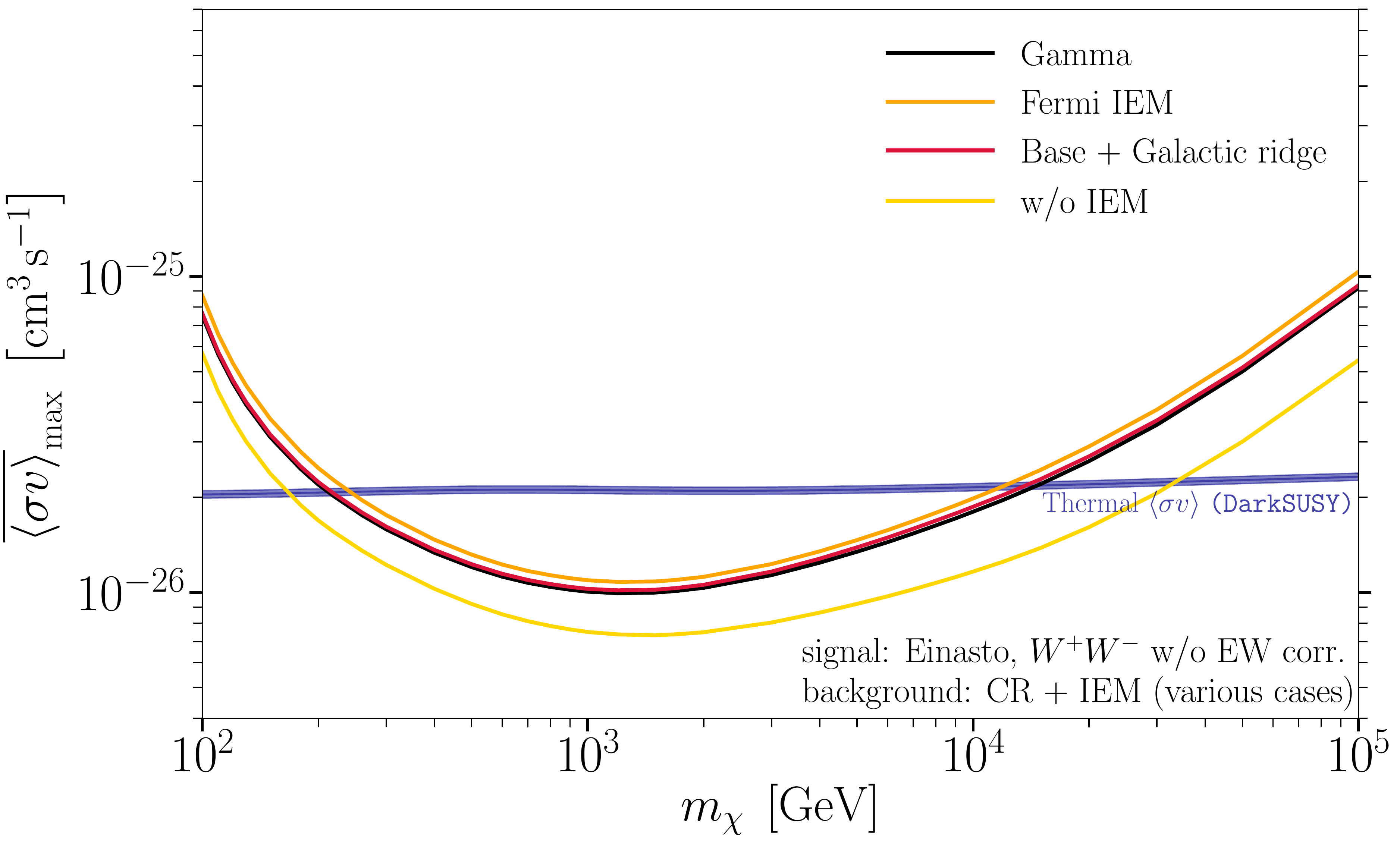}
\centering\includegraphics[width=0.3\linewidth]{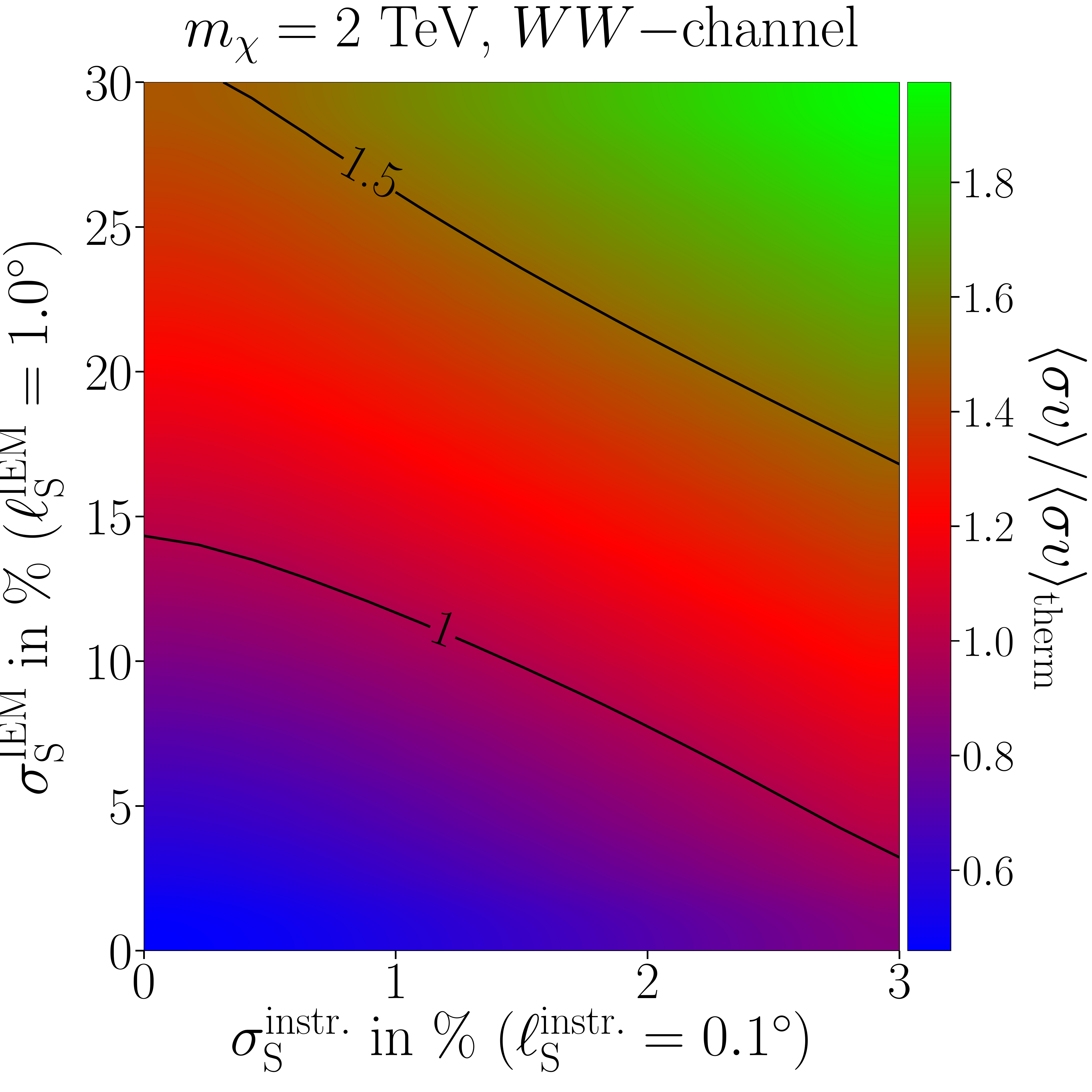}
\caption{%
{\it Left:} Projected CTA sensitivity to DM annihilation into $W^+W^-$ assuming either no IE (yellow line) or the three IE models described in Sec.~\ref{sec:astro_background}.
{\it Right:} Combined effect of instrumental and IE model systematic uncertainties on the projected DM upper limits 
(w.r.t.~the $W^+W^-$-channel, at fixed DM $m_{\chi} = 2$ TeV and for an 
Einasto profile) assuming fixed correlation lengths $\ell_{\rm S}^{\rm instr.}=0.1^\circ$ and $\ell_{\rm S}^{\rm GDE}=1.0^{\circ}$. 
The colour scheme encodes the upper limit on the annihilation cross-section in units of the thermal annihilation cross-section.
\label{fig:uncertainty_GDE}}
\end{figure}

\noindent\textbf{Impact of IE systematic uncertainties.} In Fig.~\ref{fig:uncertainty_GDE} we quantify the impact of IE model uncertainties on the project DM upper limits (for peaked DM profiles). To this end, we re-run our analysis with the remaining IE model introduced in Sec.~\ref{sec:astro_background} besides the benchmark IE choice as well as without any IE model. It follows from the left panel that the inclusion of the interstellar emission deteriorates the projected upper limits by up to a factor of 2 (c.f.~the yellow and black lines). However, the scatter of limits due to different IE models is quite weak. It must be noted, though, that this kind of analysis always assumes a `perfect' knowledge of the underlying morphology of the IE since model and mock data contain the same respective IE template. This assumption is relaxed in the right panel of this figure where we introduce a second source of systematic error, namely the model uncertainty of the IE template. It is characterised via an spatial correlation matrix with fixed correlation length $\ell_{\rm S}^{\rm GDE}=1.0^{\circ}$ but variable error amplitude. At the same time, the amplitude of the instrumental systematic uncertainty is also freed while its correlation length is fixed to the benchmark value. For an example DM particle of 2 TeV annihilating into $W^+W^-$ states, we scan the resulting upper limits (colour-coded in units of the thermal annihilation cross-section) when scanning over the displayed values for both uncertainty amplitudes. As this panel suggests, CTA will be able to probe models with only slightly enhanced annihilation rates even in the presence of instrumental systematic errors exceeding the current design goal {\it and} IE model uncertainties as large as $30\%$ (and more).

\section{Conclusions}
\label{sec:conclusions}

Our analysis has quantitatively corroborated CTA's promise to be the next-generation gamma-ray telescope being able to probe the thermal annihilation cross-section for heavy DM particles with masses in the multi-TeV range (see Fig.~\ref{fig:bb_MCband}). Even in the light of systematic errors of the instrument as well as theoretical model uncertainties of conventional astrophysics, this statement remains valid, thus rendering our sensitivity forecast robust and realistic (see Fig.~\ref{fig:uncertainty_GDE}). Together with the existing constraints on the parameter space of WIMP DM from current-generation telescopes, CTA will provide a unique opportunity to test the WIMP hypothesis and contribute to a final experimental verdict on this solution to the enigmatic nature of DM. 

\textbf{Acknowledgements.} This research has made use of the CTA instrument response functions provided by the CTA Consortium and Observatory, see \href{http://www.cta-observatory.org/science/cta-performance}{http://www.cta-observatory.org/science/cta-performance} for more details. We gratefully acknowledge financial support from the agencies and organizations listed here: \href{http://www.cta-observatory.org/consortium_acknowledgments/}{http://www.cta-observatory.org/consortium\_acknowledgments/}

\bibliographystyle{JHEP}
{\footnotesize \bibliography{CTA_GC_DM.bib}}

%
%
%
\clearpage
%
%
%
\section*{Full Authors List: The Cherenkov Telescope Array Consortium July 2021 Authors}

H.~Abdalla\textsuperscript{1}, H.~Abe\textsuperscript{2},
S.~Abe\textsuperscript{2}, A.~Abusleme\textsuperscript{3},
F.~Acero\textsuperscript{4}, A.~Acharyya\textsuperscript{5}, V.~Acín
Portella\textsuperscript{6}, K.~Ackley\textsuperscript{7},
R.~Adam\textsuperscript{8}, C.~Adams\textsuperscript{9}
% ...

1 : Centre for Space Research, North-West University, Potchefstroom, 2520, South Africa

2 : Institute for Cosmic Ray Research, University of Tokyo, 5-1-5, Kashiwa-no-ha, Kashiwa, Chiba 277-8582, Japan

3 : Pontificia Universidad Católica de Chile, Av. Libertador Bernardo O'Higgins 340, Santiago, Chile

4 : AIM, CEA, CNRS, Université Paris-Saclay, Université Paris Diderot, Sorbonne Paris Cité, CEA Paris-Saclay, IRFU/DAp, Bat 709, Orme des Merisiers, 91191 Gif-sur-Yvette, France

5 : Centre for Advanced Instrumentation, Dept. of Physics, Durham University, South Road, Durham DH1 3LE, United Kingdom

6 : Port d'Informació Científica, Edifici D, Carrer de l'Albareda, 08193 Bellaterrra (Cerdanyola del Vallès), Spain

7 : School of Physics and Astronomy, Monash University, Melbourne, Victoria 3800, Australia

8 : Laboratoire Leprince-Ringuet, École Polytechnique (UMR 7638, CNRS/IN2P3, Institut Polytechnique de Paris), 91128 Palaiseau, France

9 : Department of Physics, Columbia University, 538 West 120th Street, New York, NY 10027, USA

% ...


\end{document}